# Improving the Cauchy-Schwarz inequality

Kamal Bhattacharyya[*]

Harish-Chandra Research Institute, Chhatnag Road, Jhusi, Allahabad 211019, India

**Abstract**

We highlight overlap as one of the simplest inequalities in linear space that yields a number of useful results. One obtains the Cauchy-Schwarz inequality as a special case. More importantly, a variant of it is seen to work desirably in certain singular situations where the celebrated inequality appears to be useless. The basic tenet generates a few other interesting relations, including the improvements over certain common uncertainty bounds. Role of projection operators in modifying the Cauchy-Schwarz relation is noted. Selected applications reveal the efficacy.

[*]kamalbhattacharyya@hri.res.in

# 1. Introduction

Inequalities are encountered in many areas of theoretical sciences [1-2]. There exist quite a few standard routes [1] to arrive at specific types of inequalities. Our modest aim here is to explore how far one can extract useful results starting from a remarkably simple idea, *viz.*, the 'overlap'. To pursue, we start from the intuitively obvious inequality for two normalized states $\psi_{N1}$ and $\psi_{N2}$ in a linear space as

$$1 = \sqrt{\langle \psi_{N1} | \psi_{N1} \rangle} \sqrt{\langle \psi_{N2} | \psi_{N2} \rangle} \geq |\langle \psi_{N1} | \psi_{N2} \rangle| = S \tag{1}$$

where $S$ is the overlap. It simply means that the overlap of any two unit-norm states is never greater than unity. Inequality (1) is worst if $\langle \psi_{N1} | \psi_{N2} \rangle = 0$, and it is saturated (equality) when $\psi_{N2} = \psi_{N1}$. Thus, $S$ is a direct measure of *nonorthogonality* of the two chosen states. Sometimes, $S$ is also interpreted as 'distance' between two *pure* states, while its square is termed as *fidelity*. One may wonder that the message of (1) can be fruitfully employed to obtain the Cauchy-Schwarz inequality (CSI) and related ones, an *improved* CSI (ICSI), and sometimes a *tighter* CSI involving projection operators.

# 2. A few known results

Appropriate choices of the states in (1) may now be seen to yield certain known results, as outlined below:

**(i)** Choose two arbitrary states $\psi_1$ and $\psi_2$ such that

$$\psi_{N1} = \psi_1 / \sqrt{\langle \psi_1 | \psi_1 \rangle}, \ \psi_{N2} = \psi_2 / \sqrt{\langle \psi_2 | \psi_2 \rangle}. \tag{2}$$

Inequality (1) then quickly takes the familiar form of the Schwarz inequality [2], *viz.*,

$$\sqrt{\langle \psi_1 | \psi_1 \rangle} \sqrt{\langle \psi_2 | \psi_2 \rangle} \geq |\langle \psi_1 | \psi_2 \rangle|. \tag{3}$$

If the states $\psi_1$ and $\psi_2$ are expanded in terms of an orthonormal set of states $\{\phi_k\}$ as

$$\psi_1 = \sum_{k=1}^{N} a_k \phi_k, \ \psi_2 = \sum_{k=1}^{N} b_k \phi_k, \tag{4}$$

then (3) leads to Cauchy's inequality [2]

$$\sqrt{\sum |a_k|^2} \sqrt{\sum |b_k|^2} \geq \left|\sum (a_k^* b_k)\right|. \tag{5}$$

For real $\{a_k\}$ and $\{b_k\}$, it further simplifies in terms of average (over $N$) values as

$$\left(\overline{a^2} \ \overline{b^2}\right)^{1/2} \geq |\overline{ab}| \tag{6}$$

that possesses some relevance to statistics. Indeed, (3) and (5) are equivalent, and hence (3) is often also called the CSI. Needless to mention, the CSI (3) is usually derived from the relation $\langle \psi | \psi \rangle \geq 0$ where $\psi$ is an arbitrary normalizable state.

**(ii)** Another choice in (1), *viz.*,

$$\psi_{N1} = A^n \phi / \sqrt{\langle A^{2n} \rangle}, \ \langle A^{2n} \rangle = \langle \phi | A^{2n} | \phi \rangle \qquad (7)$$
$$\psi_{N2} = A^m \phi / \sqrt{\langle A^{2m} \rangle}, \ \langle A^{2m} \rangle = \langle \phi | A^{2m} | \phi \rangle$$

where $\phi$ is any normalized state and $A$ is hermitian, leads to

$$\sqrt{\langle A^{2m} \rangle \langle A^{2n} \rangle} \geq \left| \langle A^{m+n} \rangle \right|. \qquad (8)$$

This result primarily connects the various moments of a spatial distribution for $A = x$. However, it may also be useful elsewhere.

**(iii)** The CSI is usually employed to obtain the uncertainty product inequality (UPI). A direct application of (1), however, acts with equal facility. Define $\psi_1$ and $\psi_2$ in terms of two non-commuting hermitian operators $A$ and $B$ acting on some *arbitrary* normalized state $\phi$ as

$$\psi_1 = \phi_A = (A - \langle A \rangle I)\phi, \ \langle A \rangle = \langle \phi | A | \phi \rangle, \qquad (9)$$
$$\psi_2 = \phi_B = (B - \langle B \rangle I)\phi, \ \langle B \rangle = \langle \phi | B | \phi \rangle.$$

Then, states $\psi_{N1}$ and $\psi_{N2}$ may be taken in the forms

$$\psi_{N1} = \phi_A / \Delta A, \ \Delta A = \sqrt{\langle A^2 \rangle - \langle A \rangle^2}, \qquad (10)$$
$$\psi_{N2} = \phi_B / \Delta B, \ \Delta B = \sqrt{\langle B^2 \rangle - \langle B \rangle^2},$$

and the inequality (1) shows immediately

$$\Delta A \, \Delta B \geq \left| \langle \phi_A | \phi_B \rangle \right|. \qquad (11)$$

## 3. Some additional relations

It's now imperative to search for some more relations from (1) to justify its strength and worth further. To achieve this end, we again proceed point-wise:

### *An improved Cauchy-Schwarz inequality*

An important *special case* in the context of CSI (3) arises when

$$\langle \psi_1 | \psi_2 \rangle = 0 \qquad (12)$$

so that the right side becomes zero, rendering the celebrated inequality almost *useless*. Improvements of the CSI along various routes are available (see, *e.g.*, references [5]–[7] and those quoted therein). However, the problem with (12) does not seem to have attracted sufficient attention. Anyway, we have found it expedient [8] to tackle this problem by rewriting (1) as

$$1 = \sqrt{\langle \psi_{N1} | \psi_{N1} \rangle \langle \psi_{N2} | \psi_{N2} \rangle} \geq |\langle \psi_{N1} | \theta_{N1} \rangle||\langle \psi_{N2} | \theta_{N2} \rangle|. \tag{13}$$

Essentially, in (13), we employ two normalized *given* states $\psi_{Nj}$ and two similar *auxiliary* ones $\theta_{Nj}$, $j = 1, 2$. Auxiliary states are otherwise arbitrary, only the integrals at the right of (13) should exist. Now, following (13), we find the desired ICSI that reads as

$$\sqrt{\langle \psi_1 | \psi_1 \rangle} \sqrt{\langle \psi_2 | \psi_2 \rangle} \geq |\langle \psi_1 | \theta_{N1} \rangle||\langle \psi_2 | \theta_{N2} \rangle| \tag{14}$$

in place of (3). Note that condition (12) cannot cause any harm now, because that vulnerable inner product is avoided in (14). Further, (14) *can* reduce to (3) for the *specific choice*

$$\theta_{N1} = \psi_2 / \sqrt{\langle \psi_2 | \psi_2 \rangle}, \ \theta_{N2} = \psi_1 / \sqrt{\langle \psi_1 | \psi_1 \rangle}. \tag{15}$$

However, *other possibilities* do exist, and they can really bypass (12) to yield a non-zero right side in (14). The ICSI thus justifies its name and generality.

A point of secondary interest lies in *strengthening* the CSI when the overlap *S* in (1) is known to be *much less than* unity. Then, (3) will certainly turn out to be a *poor inequality*. Our relation (14) in such a case possesses the potential to provide better bounds.

However, while the ICSI (14) is more general than (3), it requires import of *two* auxiliary states. Elsewhere [8], we have found that this prescription too may be somewhat *relaxed* in case (12) is *exactly* valid. This new form reads as

$$\sqrt{\langle \psi_1 | \psi_1 \rangle} \sqrt{\langle \psi_2 | \psi_2 \rangle} \geq 2|\langle \psi_1 | \theta_N \rangle||\langle \psi_2 | \theta_N \rangle|; \langle \psi_1 | \psi_2 \rangle = 0. \tag{16}$$

In (16), unlike (14), just *one auxiliary state* is involved.

Finally, let us now have a look at the UPI (11). It may be beset with similar trouble as outlined above under the condition [*cf.* (9)]

$$\langle \phi_A | \phi_B \rangle = 0. \tag{17}$$

Our bypass route [8] in this case will be similar. For example, one obtains by using ICSI (14) the following result

$$\Delta A \Delta B \geq |\langle \phi_A | \theta_{N1} \rangle||\langle \phi_B | \theta_{N2} \rangle| \tag{18}$$

in place of (11). The disaster [*cf.* condition (17)] is thus avoided. This is indeed a *modified* UPI. Moreover, in view of (16), one may add an extra step to arrive at a simpler version of (18), *viz.*,

$$\Delta A \Delta B \geq 2|\langle \psi_A | \theta_N \rangle||\langle \psi_B | \theta_N \rangle| \tag{19}$$

if (17) is obeyed. The embedded *auxiliary states* in (14) or (18), or the single auxiliary state in (16) or (19), may be chosen at will, so much so that saturation can occur [8] in either case.

*A refined uncertainty sum inequality*

Kinship of the CSI with another inequality, *viz.*,

$$\sqrt{\langle \psi_1 | \psi_1 \rangle} + \sqrt{\langle \psi_2 | \psi_2 \rangle} \geq \sqrt{\langle \psi_1 - \psi_2 | \psi_1 - \psi_2 \rangle}, \tag{20}$$

is well-known [3]. We employ (20) with the choice [*cf.* definition (9)]

$$\psi_1 = \phi_A, \psi_2 = \phi_B \tag{21}$$

to obtain, for example, the result

$$\Delta A + \Delta B \geq \Delta(A - B). \tag{22}$$

On the other hand, by replacing $-\psi_2$ for $\psi_2$ in (20), the same definition (21) yields

$$\Delta A + \Delta B \geq \Delta(A + B). \tag{23}$$

The left side of (22) or (23) involves an *uncertainty sum* and hence we call such a relation as an uncertainty sum inequality (USI). Combining (22) and (23), we get

$$\Delta A + \Delta B \geq \max\{\Delta(A - B), \Delta(A - B)\}. \tag{24}$$

While inequalities like (22)-(24) may be useful, they all follow from (20). Therefore, *three weaknesses* of (20) should be pointed out here: (i) It becomes trivial when $\psi_2 = \psi_1$. (ii) Unlike (3), relation (20) is not invariant with respect to norms of $\psi_1$ and $\psi_2$. (iii) The choice (21), coupled with condition (17), leads one to an obvious result, *viz.*,

$$\Delta A + \Delta B \geq \sqrt{\Delta A^2 + \Delta B^2}. \tag{25}$$

Notably, under this situation [e.g., (17) and (21)], the right side of (25) also replaces the same of either of inequalities (22) to (24). Thus, (25) essentially relates two kinds of USI, one based on *standard deviation* and the other on *variance*. We next notice how (1) [or (13)] can save us from such inconsequentialities. Let us recall the way of construction of the ICSI (14). In the present context, we put it as

$$\sqrt{\langle \psi_1 | \psi_1 \rangle} + \sqrt{\langle \psi_2 | \psi_2 \rangle} \geq |\langle \psi_1 | \theta_{N1} \rangle| + |\langle \psi_2 | \theta_{N2} \rangle|. \tag{26}$$

This relation improves (20). None of the aforesaid weaknesses of (20) [see, *e.g.*, the discussion below (24)] now prevail. Further, using (21), an improved USI (IUSI) is found from (26), *viz.*,

$$\Delta A + \Delta B \geq |\langle \phi_A | \theta_{N1} \rangle| + |\langle \phi_B | \theta_{N2} \rangle|, \tag{27}$$

that avoids again any direct inner product between states $\phi_A$ and $\phi_B$. It is thus stronger than (24) and can even *saturate*, in contrast to (25). Only under a *specific* situation, when we choose $\theta_{N1} = \theta_{N2} = \theta_N$ in (27) and express it [remembering (17)] as a linear combination of $\phi_A$ and $\phi_B$, the *best* choice leads one to (25). However, *more general* choices exist and they do attest the generality of IUSI (27).

## *Role of projection operators*

Use of a suitable projection operator may sometime *increase the tightness* of the CSI by appropriately redressing the overlap. For simplicity, we rearrange inequality (3) in the form

$$|\langle\psi_1|\psi_2\rangle| \leq \sqrt{\langle\psi_1|\psi_1\rangle}\sqrt{\langle\psi_2|\psi_2\rangle}. \tag{28}$$

To tighten this inequality, let us keep the $\psi_2$ part as such, but incorporate an auxiliary state $\psi_3$ in the $\psi_1$ part of (28) that is known to *naturally* satisfy $\langle\psi_2|\psi_3\rangle=0$. Such a choice renders the left side unaltered, though form (28) changes to

$$|\langle\psi_1+\alpha\psi_3|\psi_2\rangle| = |\langle\psi_1|\psi_2\rangle| \leq \sqrt{\langle\psi_1+\alpha\psi_3|\psi_1+\alpha\psi_3\rangle}\sqrt{\langle\psi_2|\psi_2\rangle}. \tag{29}$$

Thus, value of the right side changes. A little algebra shows that the tightest situation, *i.e.*, *minimum value for the right side* in (29), is attained at an optimum $\alpha$ to yield from (29)

$$|\langle\psi_1|\psi_2\rangle| \leq \sqrt{\langle\psi_1|\psi_1\rangle - \frac{|\langle\psi_1|\psi_3\rangle|^2}{\langle\psi_3|\psi_3\rangle}}\sqrt{\langle\psi_2|\psi_2\rangle}. \tag{30}$$

It implies, we can reorganize (29) in the *tightest situation* as

$$|\langle\psi_1|\psi_2\rangle| \leq \sqrt{\langle(I-P_3)\psi_1|(I-P_3)\psi_1\rangle}\sqrt{\langle\psi_2|\psi_2\rangle}. \tag{31}$$

In (31), $P_3$ refers to the projector for $\psi_3$ and is defined by

$$P_3 = |\psi_{N3}\rangle\langle\psi_{N3}|; \psi_{N3} = \psi_3/\sqrt{\langle\psi_3|\psi_3\rangle}. \tag{32}$$

Thus, (28) admits modification to (31) when a state $\psi_3$ is known *a priori* to be manifestly orthogonal to $\psi_2$. In our form (14) or (16), on the contrary, the only restriction would be $\theta_{N2}, \theta_N \neq \psi_3$. Of course, when the whole of $\psi_1$ is orthogonal to $\psi_2$, the left side in (31) needs to be rectified, and there appears our prescription, the ICSI (14) or (16), as a remedy.

In the context of UPI [recall (9)], this projector issue is nicely met in the standard route. By construction, $\langle\phi_B|\phi\rangle=0$, and so we choose $\phi_A$ such that $\langle\phi_A|\phi\rangle=0$. Indeed, this is in-built in the definition. Had we chosen instead, *e.g.*, $\phi'_A = A\phi$, we would have been led to a weaker inequality [as in (28)], keeping aside the fact that this option *does not* involve the standard deviation of operator *A*. However, by following (31), one regains the usual form, as found in (9). Therefore, while applying (1), we shall continue with the above wisdom in studies on the complementary Eckart bound [9] and survival probability [10-13] to follow.

## 4. Results and applications

Let us quickly turn to certain results that will demonstrate the advantage of the present endeavor.

<u>CSI vs. ICSI</u>

We take the following states for a first-hand experience:

$$\psi_{N1} = \tfrac{1}{\sqrt{1+x^2}}(x \ \ 1)^T, \ \psi_{N2} = (1 \ \ 0)^T, \ \theta_{N1} = \tfrac{1}{\sqrt{2}}(1 \ \ 1)^T, \ \theta_{N2} = \tfrac{1}{\sqrt{1+x^2}}(1 \ \ x)^T. \tag{33}$$

One obtains the results shown in Figure 1 below for varying $x$-values. Here, the left side is fixed at unity. The right sides stand for *lower bounds* to the same. In the figure, we exhibit respectively the CSI (3) [black: 1], the ICSI (14) [red: 2], (14) with exchanging $\theta_{N1}$ and $\theta_{N2}$ [blue: 3], (14) with $\theta_{N2} = \theta_{N1}$ [magenta: 4] and (14) with $\theta_{N1} = \theta_{N2}$ [dark cyan: 5]. Note that the CSI performs nicely for large $|x|$. But, overlap is small in regions around $x = 0$ [$S = 0$ at $x = 0$], and that is the *primary focus* of the present study. We witness here varying performances of the other choices based on (14). Particularly interesting ones are curves 2 and 4. Both fare well around the $x = 0$ region. Curves 3 and 5 also perform better than curve 1 within $|x| = 1$. The advantage of using auxiliary states in (14) should now be clear. While none of the curves reach the *exact* value, we happily note that (16) applies at $x = 0$, and curve 4 shows its *ability to saturate* at this point.

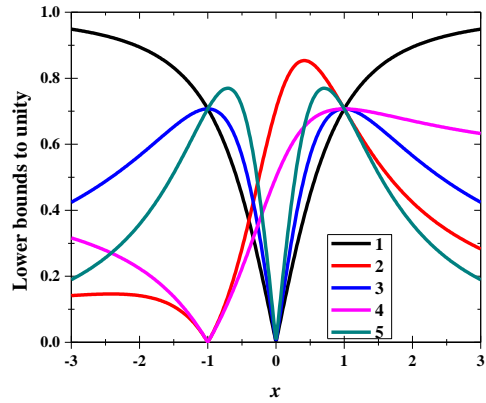

**Figure 1**

Our next example concerns *two lowest* normalized energy eigenstates of the particle-in-a-box problem in $(0, \pi)$ for which the overlap is zero, and hence the CSI (3) [$1 \geq 0$] is of no use. In this situation, we see how the ICSI (16) performs, aided by just *one auxiliary state* taken as $\theta_N = N x^2 (\pi - x)$, and zero otherwise, where $N$ stands for the normalization constant. Our recipe betters the bound from $1 \geq 0$ to $1 \geq 0.6553$. The tightness achieved is again noteworthy.

In both the above cases, however, auxiliary states of our choice are employed. One may wonder whether betterment can be accomplished at all by using *only the parent states* [*e.g.*, $\psi_{N1}, \psi_{N2}$]. To explore, we proceed by choosing

$$\theta_{N1} = N_1(c_1\psi_{N1} + c_2\psi_{N2}), \theta_{N2} = N_2(d_1\psi_{N2} + d_2\psi_{N1}). \tag{34}$$

Let us also take all the states as real, with real positive combining coefficients and overlap. The prefactors $N_1$ and $N_2$ in (34) represent the normalization constants. Figure 2 attests that *the whole accessible region* at $S = 0$ is better in ICSI (14). Even, saturation is possible here. Generality of the ICSI may now be appreciated.

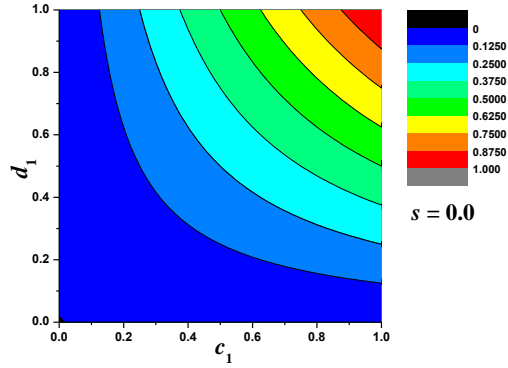

**Figure 2**

A complementary Eckart bound

The Eckart inequality [9] stands as one of the earliest measures for the goodness of an approximate normalized eigenstate of energy $\bar{\phi}_1$ by providing a *lower bound* to its overlap with the unknown exact *ground* stationary state $\phi_1$. Writing the energy eigenvalue equation as $H\phi_j = \varepsilon_j\phi_j, \|\phi_j\| = 1$, and defining the average ground-state energy as $\bar{\varepsilon}_1$, one finds [9] that

$$S_1^2 = \left|\langle \bar{\phi}_1 | \phi_1 \rangle\right|^2 \geq \frac{\varepsilon_2 - \bar{\varepsilon}_1}{\varepsilon_2 - \varepsilon_1}; \bar{\varepsilon}_1 = \langle \bar{\phi}_1 | H | \bar{\phi}_1 \rangle. \tag{35}$$

Coupled with the upper bound $S_1 \leq 1$, the above result actually reflects the closeness of $\bar{\phi}_1$ with the *unknown* $\phi_1$. However, an improved upper bound may be obtained by defining

$$\begin{aligned}\psi_1 &= (A - \langle A \rangle I)\bar{\phi}_n, \ A = |\phi_n\rangle\langle\phi_n|, \langle A \rangle = \langle \bar{\phi}_n | A | \bar{\phi}_n \rangle = S_n^2, \Delta A = S_n\sqrt{1-S_n^2}, \\ \psi_2 &= (H - \langle H \rangle I)\bar{\phi}_n, \langle H \rangle = \langle \bar{\phi}_n | H | \bar{\phi}_n \rangle = \bar{\varepsilon}_n, \Delta H = \Delta\varepsilon_n, S_n = |\langle \bar{\phi}_n | \phi_n \rangle|, \end{aligned} \tag{36}$$

and subsequently using (1). The outcome is

$$\frac{S_n}{\sqrt{1-S_n^2}} \leq \frac{\Delta\varepsilon_n}{|\bar{\varepsilon}_n - \varepsilon_n|}. \tag{37}$$

This is *complementary* to the standard Eckart bound. Moreover, unlike (35), (37) applies to any *n*-th state, not *just* the *ground* state, and it does not require any information about $\varepsilon_2$. Our preliminary checks reveal that (37) furnishes far better bound than the primitive one, *viz.*, $S_n \leq 1$, primarily due to the denominator at the left. For any finite right side in (37), $S_n$ at the left *has to be less* than unity. Had we chosen $\psi_1 = A\bar{\phi}_n$ in (36) instead, we would not reach this strong form.

<u>Decay probability</u>

Turning to quantum dynamics, we now consider the problem of survival probability $P(t)$ [10-13] or, more specifically, the decay probability $Q(t)$. To proceed, we first define a state $\psi(t)$ whose evolution is governed by a conservative Hamiltonian $H$ as

$$\psi(t) = \exp(-iHt/\hbar)\psi(0); \|\psi(0)\| = 1. \tag{38}$$

Next, we identify below a specific projection operator $A$ whose average in state $\psi(t)$ yields $P(t)$.

$$A = |\psi(0)\rangle\langle\psi(0)|; \langle A\rangle = \langle\psi(t)|A|\psi(t)\rangle = P(t) \tag{39}$$

The states in (1) are now chosen in the forms [*cf.* (9) and (10)]

$$\psi_{N1} = (A - \langle A\rangle I)\psi(t)/\sqrt{P(t)\cdot Q(t)}, Q(t) = 1 - P(t);$$
$$\psi_{N2} = (H - \langle H\rangle I)\psi(t)/\Delta E; \Delta E^2 = \langle\psi(t)|(H - \langle H\rangle I)^2|\psi(t)\rangle. \tag{40}$$

Putting these in (1), we find after a little algebra the inequality

$$\Delta E\sqrt{Q(t)} \geq |\langle\psi(0)|H - \langle H\rangle I|\psi(t)\rangle|. \tag{41}$$

But, a direct application of the CSI (3) for the right side of (41) leads us to a *weaker bound*, *viz.*,

$$|\langle\psi(0)|H - \langle H\rangle|\psi(t)\rangle| \leq \Delta E. \tag{42}$$

Notice, the multiplying factor $Q(t)$ [$Q(t) \leq 1$] at the left of (41) does not appear at the right side of (42), and so the latter loses the time dependence. Indeed, this becomes decisive in *tightening* the bound (41) compared to (42), chiefly at *short times* when the starting state changes little, and hence $Q(t) \ll 1$ follows. At $t = 0$, however, both the left and right sides of (41) are zero; hence, there is no paradox. Thus, we again witness the worth of (1). Role of the projector is also evident in the choice (40).

The next task would naturally be to put (41) to test. In the short-time regime, one obtains from (41) a general result of the form

$$\sqrt{Q(t)} \geq \frac{\Delta E t}{\hbar}\left[1 + \frac{t^2\left(\langle H^3\rangle - \langle H^2\rangle\langle H\rangle\right)^2}{8\hbar^2 \Delta E^4} + ...\right]. \tag{43}$$

However, depending on the eigenvalue spectrum of $H$ in (38), two situations should now be distinguished. If the spectrum is continuous, the state gradually decays. In contrast, one observes decay and revival in succession when $H$ has a purely discrete spectrum. This quantum recurrence [14-16] is important in various areas [17]. We consider the efficacy of (41) in both these cases.

Concentrating first on pure decay [12], an upper bound to the decay probability $Q(t)$ [see Eq. (10) in Ref. 12] may be found as

$$\sqrt{Q(t)} \leq \sin(\Delta E t / \hbar),\ 0 \leq t \leq \tfrac{\pi\hbar}{2\Delta E}. \tag{44}$$

In (41), however, we have arrived at a *complementary bound*. What is more, while (44) is valid only over short times, our present lower bound (41) does not impose any such restriction on time. As an example, let us pay attention to a solvable problem, *viz.*, the decay of a Gaussian packet [3, 12] in field-free space. Implementing (41), we observe after some algebra that the standard energy form factor [12] finally yields

$$\sqrt{Q(t)} \geq \frac{(\Delta E t / \hbar)}{\left(1 + 2(\Delta E t / \hbar)^2\right)^{3/4}}. \tag{45}$$

We have checked that this relation is valid over the entire region of time. But, as already stated, inequality (41) works *better* at smaller $t$. Thus, at a time when $\Delta E t / \hbar = 1/4$, one finds from (45) $\sqrt{Q(t)} \geq 0.23$, whereas (44) gives $\sqrt{Q(t)} \leq 0.25$. The exact result [12] yields $\sqrt{Q(t)} \approx 0.24$, justifying the tightness of either bound.

Let us next focus on quantum recurrences. A preliminary check reveals that (41) is *exact* for any 2-level problem. This is specifically comforting in view of its direct relevance with *quantum speed limits* [18 - 20]. To be explicit, choosing $H\phi_j = \hbar\omega_j \phi_j, \|\phi_j\| = 1,$ and $j = 1, 2, ...,$ we obtain

$$\sqrt{Q(t)} = 2r_1 r_2 |\sin(\omega_{21} t / 2)|,\ \omega_{21} = \omega_2 - \omega_1, |c_j| = r_j,$$
$$r_1^2 + r_2^2 = 1, \Delta E = r_1 r_2 \hbar \omega_{21}, \langle H \rangle = \hbar(r_1^2 \omega_1 + r_2^2 \omega_2) \tag{46}$$

where

$$\psi(0) = \sum_{j=1}^{2} c_j \phi_j. \tag{47}$$

Choosing $r_1 = \cos\theta$ in (46), one finds

$$\sqrt{Q(t)} = \sin 2\theta |\sin(\omega_{21} t / 2)|, \quad \Delta E = \tfrac{1}{2} \hbar \omega_{21} \sin 2\theta. \tag{48}$$

A few remarks are now in order. First, recurrences begin with decay, and we shall consider this *primary decay part* below. Secondly, (46) or (48) shows a certain *symmetry* with respect to *exchange* of $r_1$ and $r_2$ for both $Q(t)$ and $\Delta E$, something that is *lacking* in the average energy $\langle H \rangle$. Thirdly, from (43) and (44), we see that a $t$-$\sqrt{Q(t)}$ plot is initially linear. Indeed, one obtains

$$\lim_{t \to 0} \Delta E t = \hbar \sqrt{Q(t)}. \tag{49}$$

Here, $t$ is the dynamical time. Relation (49) ties the energy uncertainty with the decay probability and time, and applies to *both decay and recurrence* problems. Fourthly, (48) shows that the $\theta = \pi/4$ case (*equiprobable*) decays *most rapidly* to $Q(t) = 1$ (the orthogonal state) at $t = \pi/\omega_{21}$. Calling this time as $\tau$, we obtain $\tau = h/(4\Delta E)$ using (48). This is the Fleming bound [11] for quantum speed [18 - 20]. If $\omega_1 = 0$, one also finds the Margolus-Levitin bound [21], *viz.*, $\tau = h/(4\langle H \rangle)$ from (46) and (48). Fifthly, the maximum decay for any *general, non-equiprobable* situation is given by $\sqrt{Q(t)} = \sin 2\varphi, \varphi \neq \pi/4$. This is also reached at $t = \pi/\omega_{21} = \tau_g$ (say). On the other hand, *along the fastest decay route* (equiprobable), the result $\sqrt{Q(t)} = \sin 2\varphi$ is achieved at a time $\tau_e = 4\varphi/\omega_{21}$. These results along with their corresponding $\Delta E$-values are summarized below.

$$\theta = \varphi \neq \pi/4: \sqrt{Q(t)_g^{\max}} = \sin 2\varphi, \ \tau_g = \pi/\omega_{21}, \ \Delta E_g = \frac{\omega_{21}}{2} \sin 2\varphi$$
$$\theta = \pi/4: \sqrt{Q(t)_e} = \sin 2\varphi, \ \tau_e = 4\varphi/\omega_{21}, \ \Delta E_e = \frac{\omega_{21}}{2} \tag{50}$$

One observes now that the relative time and the relative energy spread obey

$$\tau_{rel} = \tau_g / \tau_e = \pi/4\varphi, \ \Delta E_{rel} = \Delta E_g / \Delta E_e = \sin 2\varphi. \tag{51}$$

For small enough $\varphi$, (51) shows that $\tau_{rel}$ becomes *exceedingly large*. The dependence of the relative decay time on the initial state of a qubit should now be transparent. However, there exists a nice equality in this regime connecting $\tau_{rel}$ and $\Delta E_{rel}$, *viz.*,

$$\tau_{rel} \Delta E_{rel} = \pi/2 \left[1 + O(\varphi^2)\right], \ \varphi \to 0. \tag{52}$$

In (52), *h* does not appear at the right just because each term at the left is dimensionless.

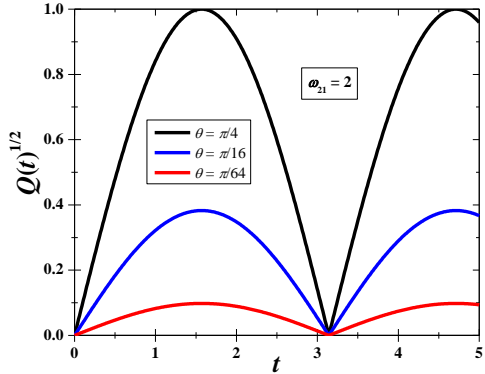

**Figure 3**

Figure 3 displays how the maximum of $Q(t)$ reduces with decreasing contribution of one of the two states, denoted here by $\theta$. A concomitant increase in the minimum time to attain some pre-assigned value of $Q(t)$ is also clear. In these situations, the *minimum-time bounds* set by the fastest decay route could be quite *useless*. One may instead concentrate on (52). The figure clarifies the critical role of the *initial state*, as emphasized elsewhere [22].

In fine, we also note that the Fleming bound applies to both pure decay and recurrence problems, but the Margolus-Levitin bound concerns the latter situation only.

## 5. Conclusions

To summarize, we sketched here how the overlap inequality (1) can be exploited to obtain the CSI (3). More importantly, in case the CSI fails to work [*e.g.*, under condition (12)], a form of the overlap inequality (13) leads to an ICSI (14) that applies to any arbitrary situation. A further simplification to ICSI (16) is a positive addition. The UPI (18) or (19) reveals the gains in suitable contexts. For the USI case, a similar extension of (14) to (26) yields an IUSI (27) that is more general than the prevalent form. We explored also the worth of a projector in tightening the CSI [*cf.*, relations (28) - (32)].

The ICSI (14) and (16) are applied to a few pathological situations in Figures 1 and 2. These are general linear-space problems. Auxiliary states are imported in Figure 1 to specifically study the $x \rightarrow 0$ limit. However, in Figure 2, such states are avoided. The benefit of a projector is highlighted in certain areas of quantum mechanics. These studies include a complementary Eckart bound, bounds to the decay probability $Q(t)$, and a few other characteristics of the latter.

**References**

1. G. H. Hardy, J. E. Littlewood, G. Polya, *Inequalities* (CUP, 1952).

2. M. Abramowitz, I. A. Stegun, Eds. *Handbook of Mathematical Functions*, Ch. 3 (Dover, NY, 1965).

3. J. L. Powell, B. Crasemann, *Quantum Mechanics*, (Addision Wesley, Reading, Mass., 1961).

4. E. Merzbacher, *Quantum Mechanics*, (John Wiley, NY, 1961)

5. C. Lupu, D. Schwarz, Appl. Math. Comput. **231**, 463 (2014).

6. I. Pinelis, Am. Math. Mon. **122**, 593 (2015).

7. S. G. Walker, Stat. Prob. Lett. **122**, 86 (2017).

8. K. Bhattacharyya, arXiv: 1607.07331.

9. C. Eckart, Phys. Rev. **36**, 878 (1930); L. Pauling, E. B. Wilson, Introduction to Quantum Mechanics (McGraw-Hill, Kogakusha, 1935).

10. L. Fonda, G. C. Ghirardi, A. Rimini, Rep. Prog. Phys. **41**, 587 (1978).

11. G. N. Fleming, Nuovo Cimento A **16**, 232 (1973).

12. K. Bhattacharyya, J. Phys. A **16**, 2993 (1983).

13. P. Pfeifer, Phys. Rev. Lett. **70**, 3365 (1993); P. Pfeifer and J. Fröhlich, Rev. Mod. Phys. **67**, 759 (1995).

14. P. Bocchieri, A. Loinger, Phys. Rev. **107**, 337 (1957).

15. A. Peres, Phys. Rev. Lett. **49**, 1118 (1982).

16. K. Bhattacharyya, D. Mukherjee, J. Chem. Phys. **84**, 3212 (1986).

17. C. Gogolin, J. Eisert, Rep. Prog. Phys. **79**, 056001 (2016).

18. V. Giovannetti, S. Lloyd, L. Maccone, Phys. Rev. A **67**, 052109 (2003).

19. M. G. Bason, M. Viteau, N. Malossi, P. Huillery, E. Arimondo, D. Ciampini, R. Fazio, V. Giovannetti, R. Mannella, O. Morsch, Nature Phys. **8**, 147 (2012).

20. O. Andersson, H. Heydari, J. Phys. A **47**, 215301 (2014)

21. N. Margolus, L. B. Levitin, Physica D **120**, 188 (1998).

22. S. Wu, Y. Zhang, C. Yu, H. Song, J. Phys. A **48**, 045301 (2015).